\documentclass[prb, preprint, letter, superscriptaddress]{revtex4}
\usepackage{amssymb}
\usepackage{amsmath}
\usepackage{bbm}
\usepackage{graphicx}
\usepackage{color}

\newcommand{\at}{\tilde\alpha}
\newcommand{\wi}{\vec{w}_i}
\newcommand{\mi}{\vec{m}_i}

\begin{document}

\author{Shenshen Wang}
\author{Peter G. Wolynes*}
\affiliation{Department of Physics, Department of Chemistry and Biochemistry,  and Center for Theoretical Biological Physics, University of California, San Diego, La Jolla, CA 92093, USA}

\title{On the spontaneous collective motion of active matter}
\date{\today}

\begin{abstract}

Spontaneous directed motion, a hallmark of cell biology, is unusual in classical statistical physics. Here we study, using both numerical and analytical methods, organized motion in models of the cytoskeleton in which constituents are driven by energy-consuming motors. While systems driven by small-step motors are described by an effective temperature and are thus quiescent, at higher order in step size, both homogeneous and inhomogeneous, flowing and oscillating behavior emerges. Motors that respond with a negative susceptibility to imposed forces lead to an apparent negative-temperature system in which beautiful structures form resembling the asters seen in cell division.

\end{abstract}

\hyphenation{}

\maketitle

Spontaneous directed motion driven by active processes is crucial to biology. Such motion is only possible because the cell is a far-from-equilibrium many-body system. The cytoskeleton of eukaryotic cells is built, maintained and adaptively reorganized through active transport and force generation powered by ATP hydrolysis. Oscillations of the mitotic spindle during cell division \cite{spindle oscillation} and cytoplasmic streaming \cite{cytoplasmic streaming} dramatically illustrate that the cell is not at equilibrium. Driven motions of cells are also important
at higher levels of organization in living things ranging from mechanosensation \cite{hair bundle oscillation} to the developmental processes in which the genetic code unfolds to create a multicellar organism \cite{embryonic development}.
Sustained spontaneous collective motion is quite remarkable in many-body physics. Superfluidity and superconductivity are examples of metastable states of motion made possible by quantum statistics. The biological example provided by the cytoskeleton is seemingly quite different leading not to infinitely long-lived states but to ones that go away when the cell is depleted of fuel and dies. Nevertheless, like the quantum examples, the motion of the cytoskeleton is an emergent many-body phenomenon reflecting broken symmetries.

Here we explore the origin of spontaneous collective motion for systems of many interacting biomacromolecules with motor-driven active processes using a systematic perturbative expansion of the many-body master equation treating nonequilibrium motorized processes.
We model the motors as generating a time series of isotropic kicks on the constituents of a many-body assembly. Earlier \cite{Teff ppr} we showed that quite generally the corresponding master equation, when expanded to the lowest order in the kick step size, yields an effective temperature, $T_\textrm{eff}$, which explicitly depends on the total motor activity and on the way in which motors respond to imposed forces. A system described by an effective temperature alone \cite{Teff theory1, Teff theory2, Teff theory3} cannot undergo spontaneous directed motion unless it is quantum mechanical.
Pursuing the expansion to higher order, however, reveals the possible emergence of spontaneous directed collective motion quite generally from a quiescent homogeneous state, albeit one with rigidity owing to broken translational symmetry, as in a glass. The underlying dynamic instability is induced by a sufficiently strong internal agitation in terms of kick step size. This provides a general mechanism for spontaneous flows in an active assembly of interacting constituents.

Combining a linear stability analysis with a trial solution of the many-body master equation allows us to identify possible dynamic phases that depend on the motor kick step size and susceptibility. We find that for sufficiently large kicks and high activity, susceptible motors, i.e., motors whose kick rate depends on the forces exerted on them, can generate spontaneous flow, whereas adamant motors, indifferent to imposed forces, would merely drive fluidization of an active system. We have also carried out simulations on a minimal cytoskeleton model incorporating motor dynamics to compare with our analytical predictions. The simulations not only verify the predicted phase diagram, but also highlight how the combination of network connectivity with motor susceptibility determines the formation of nonequilibrium structures. The simulations show an oscillatory phase separation at intermediate network connectivity and formation of aster-like patterns/bundle-connected poles when driven by motors with negative susceptibility, i.e., motors that move against the force, energetically uphill. The latter corresponds to a negative-temperature system where interesting structures emerge much like vortex condensation in two-dimensional turbulence \cite{vortex condensation, turbulent flow}.

We are far from the first to try to understand the physics of spontaneous collective motion in biology.
J\"{u}licher and Prost \cite{spontaneous motion_Julicher} studied a one-dimensional stochastic model which assumed an underlying ratchet potential already breaking translational symmetry. Motor cooperativity then leads to a dynamical phase transition to spontaneous directed motion despite the system's spatial symmetry.
Thinking of the cytoskeleton, an assembly of filamentous polar polymers actively connected by crosslinkers, as an active polar gel has allowed the construction of continuum theories, based on conservation laws and symmetry considerations, which also generate active flows \cite{generic phase diagram, confinement, HD_solution, HD_gel}. Pattern formation in active fluids has also been discussed based on a reaction-diffusion-advection mechanism \cite{reaction_diffusion_advection}.

Here we model the stochastic nature of the motor kicking via a master equation for the many-body probability distribution function $\Psi(\{\vec{r}\},t)$ \cite{TY1, TY2}
\begin{equation}
\frac{\partial}{\partial t}\Psi(\{\vec{r}\},t)=(\hat{L}_\textrm{FP}+\hat{L}_\textrm{NE})\Psi(\{\vec r\},t).
\end{equation}
Here $\hat{L}_\textrm{FP}=D_0\sum_i\nabla_i\cdot\nabla_i-D_0\beta\sum_i\nabla_i\cdot(-\nabla_i U)$
is the usual many-body Fokker-Planck operator describing passive Brownian motion with $D_0$ denoting the ordinary diffusion coefficient at ambient temperature $T$ and $\beta=1/k_\textrm B T$. The gradients of the many-body interaction free energy $U(\{\vec r\})=U(\vec{r}_1,\vec{r}_2,\cdots,\vec{r}_n)=\sum_{<ij>}u(\vec{r}_{i}-\vec{r}_j)$ give the local forces acting on individual particles, where $\vec{r}_i$ is the position of the $i$th particle and $\langle\cdots\rangle$ denotes the nearest neighbor pairs.
The effects due to nonequilibrium motorized processes are summarized by an integral kernel
$\hat{L}_\textrm{NE}\Psi(\{\vec r\},t)=\int\Pi_i d\vec{r'_i}[K(\{\vec{r'}\}\rightarrow\{\vec{r}\})\Psi(\{\vec{r'}\},t)-K(\{\vec{r}\}\rightarrow\{\vec{r'}\})\Psi(\{\vec{r}\},t)]
$, where $K(\{\vec{r'}\}\rightarrow\{\vec{r}\})$ encodes the probability of transitions between different particle configurations per unit time.
Motor kicking noise is a finite jump process with a rate that depends on whether the free energy is increased or decreased when a step is made
\begin{equation}\label{kinetic rate}
k=\kappa[\Theta(\Delta U)\exp(-s_u\beta\Delta U)+\Theta(-\Delta U)\exp(-s_d\beta\Delta U)].
\end{equation}
Here $\Theta$ is the Heaviside step function and $\Delta U=U\left(\vec{r}+\vec{l}\right)-U\left(\vec{r}\right)$ is the energy change due to the kick identified by a vector $\vec l=l\hat n$. The kick step size $l$ and the basal kicking rate $\kappa$ define the dimensionless motor activity $\Delta:=\kappa\l^2/D_0$, an analog of the Peclet number in turbulent diffusion.
This model rate couples the chemical reactions leading to the motor activity to the local mechanical forces acting on the motor being parametrized by the susceptibility $s$ which may take different values for uphill ($s_u$) moves and for downhill ($s_d$) moves depending on the biochemical mechanism of the motors. When $s\rightarrow1$ the motors are susceptible, slowing down when they climb up against obstacles and accelerating when they move energetically downhill; in contrast $s\rightarrow0$ corresponds to completely adamant motors which kick at a rate unperturbed by the free energy landscape.

To examine the small kick limit, we first expand the equation in powers of $\vec l$ up to the quadratic order.
The simplest case, isotropic kicking and symmetric susceptibility (i.e., $s_u=s_d=s$) leads directly to an effective Fokker-Planck equation \cite{Teff ppr}
\begin{equation}\label{detailed balance}
\frac{\partial}{\partial t}\Psi(\{\vec r\},t)=D_\textrm{eff}\sum_i\Big\{\nabla_i^2\Psi-\nabla_i\cdot\left[(-\nabla_i \beta_\textrm{eff}U)\Psi\right]\Big\},
\end{equation}
where
\begin{equation}\label{Deff}
D_\textrm{eff}=D_0\left(1+\frac{1}{2d}\frac{\kappa l^2}{D_0}\right),
\end{equation}
\begin{equation}\label{Teff}
\left(\beta_\textrm{eff}/\beta\right)^{-1}=T_\textrm{eff}/T=\left(1+\frac{1}{2d}\frac{\kappa l^2}{D_0}\right)\Big/\left(1+\frac{s}{d}\frac{\kappa l^2}{D_0}\right).
\end{equation}
These simple expressions (\ref{detailed balance})--(\ref{Teff}) valid for general spatial dimensions $d$ have nontrivial implications.
In the small kick limit, the active system, while out of equilibrium, behaves as if it is at an effective canonical equilibrium characterized by an effective temperature $T_\textrm{eff}$. The effective diffusion constant $D_\textrm{eff}$ (equation~\ref{Deff}) is enhanced by the active processes regardless of motor adamancy, consistent with recent observations of enhanced cytoplasmic diffusion \cite{cytoplasmic diffusion}.
$T_\textrm{eff}$ (equation~\ref{Teff}) is fully determined by the motor activity $\Delta=\kappa l^2/D_0$ and the motor susceptibility $s$; susceptible motors with $s>1/2$ yield $T_\textrm{eff}<T$. When motor activity dominates over thermal noise, i.e., $\Delta\gg1$, the effective temperature diverges as $T_\textrm{eff}/T\sim 1/(2s)$ as $s\rightarrow0$. Thus intense kicking by adamant motors leads to a very high effective temperature just as observed in experiments \cite{Mizuno_science} and simulation studies \cite{rheological and structural, Teff of active matter}.
A more detailed discussion can be found in a separate work \cite{Teff ppr}.

To probe the dynamic instability that may give rise to the spontaneous motion, we must go beyond the effective equilibrium and expand to quartic order of the kick step size obtaining
\begin{eqnarray}\label{quartic order}
\frac{\partial}{\partial t}\Psi(\{\vec r\},t)&=&D_\textrm{eff}\sum_i\Big\{\nabla_i^2\Psi-\nabla_i\cdot\left[(-\nabla_i \beta_\textrm{eff}U)\Psi\right]\Big\}\nonumber\\
&+&\kappa l^4\langle\cos^4\theta\rangle_{\widehat n}\times\sum_iF_i(\nabla^{(m)}_iU,\nabla^{(n)}_i\Psi).
\end{eqnarray}
The complicated functional $F_i$ is the divergence of a flux,
i.e., $F_i=-\nabla_i\cdot\vec{J^a_i}$, where $\vec{J^a_i}$ is the probability current due to active events on particle $i$ given by
\begin{eqnarray}\label{Ji}
-\vec{J^a_i}&=&\frac{1}{24}\nabla_i^3\Psi+\frac{s}{12}\nabla_i\left(\nabla_i^2\beta U\Psi\right)
+\frac{s}{6}\nabla_i\beta U\nabla_i^2\Psi\nonumber\\
&+&\frac{s^2}{4}(\nabla_i\beta U)^2\nabla_i\Psi
+\frac{s^3}{6}(\nabla_i\beta U)^3\Psi.
\end{eqnarray}
Therefore while at quadratic order in $l$ a motor-driven system exhibits enhanced diffusive dynamics at an effective equilibrium, at quartic order, a net streaming flow becomes possible.

In an early study of the stability and dynamics of a motorized assembly Shen and Wolynes \cite{TY1} pictured the motors as introducing a modification to the Debye-Waller factors of the localized particles.
They found an expression of the deviation of the total localization strength $\at$ from its thermal value $\alpha$ in terms of the motor properties. Thermal self-consistent phonon theory \cite{fixman} gives $\alpha$ for a central particle that depends on the $\at$ of all its neighbors. Combining these two aspects allows a self-consistent determination of mean-field ($\alpha, \at$) solutions allowing an identification of stability limits.

Assuming $s_u=s_d=s$, the second moment closure \cite{TY2} reduces to a simple expression
$(\at-\alpha)/\at=\left(s-1/2\right)\,\exp\left[s(s-1)\alpha l^2\right]\kappa l^2/dD_0$.
Thus for $s=1/2$ ($T_\textrm{eff}=T$) chemical noise does not modify the mechanical stability ($\tilde\alpha=\alpha$); for $s<1/2$ ($T_\textrm{eff}>T$) stability is weakened ($\tilde\alpha<\alpha$) whereas for $s>1/2$ ($T_\textrm{eff}<T$) stability is enhanced ($\tilde\alpha>\alpha$).

For spontaneous collective motion to occur there must be a nontrivial dynamic first moment that indicates a moving fiducial lattice. The second moment still has its steady-state value describing vibrations about the fiducial configuration. We thus write down a trial function of the master equation as a collection of Gaussians with moving centers and a steady variance
\begin{equation}\label{Gaussian ansatz}
\Psi(\{\vec{r}_i\};\tilde\alpha)=\Pi_i\left(\tilde\alpha/\pi\right)^{d/2}e^{-\tilde\alpha[\vec{w}_i-\vec{m}_i(t)]^2},
\end{equation}
where $\vec{w}_i=\vec{r}_i-\vec{R}_i$ denotes the displacement of particle $i$ from its equilibrium position $\vec{R}_i$, $\vec{m}_i(t)=\langle\vec{w}_i|\Psi\rangle$ defines the dynamic first moment of particle $i$,
and the total localization strength $\at$ of individual particles is inversely related to the second moment.
Applying the first moment closure $\partial_t\langle\wi|\Psi\rangle=\langle\wi|(\hat{L}_\textrm{FP}+\hat{L}_\textrm{NE})\Psi\rangle$ to equation (\ref{quartic order}) with this ansatz (\ref{Gaussian ansatz}) as well as the coupled-oscillator expansion of the effective potential \cite{coupled oscillator} leads to coupled equations for the $\mi$'s.

To investigate the spontaneous emergence of directed motion from a quiescent homogeneous state, we carry out a linear stability analysis on the non-moving state (i.e. $\mi=0$).
In view of the biological relevance of collective motion in one-dimensional scenarios, such as the filament sliding in motility assay \cite{motility assay} and flow in the cell cortex \cite{cortex flow},
we first demonstrate the instability onset for the $1$D case. Detailed analysis for general dimensions will be given later.
Consider a spatially varying trial solution of the form
\begin{equation}\label{perturbation}
m_i=\bar{m}\,e^{r_kt+ikR_i},
\end{equation}
where $\bar{m}$ denotes the amplitude of the first moment, $k$ the wavenumber of the spatial modulation of a collective mode and $r_k$ the growth rate of the $k$-mode.
Even if kicks are isotropic and the interaction Hamiltonian preserves rotational symmetry, spontaneous symmetry breaking occurs giving flow in a specific direction.

The linearized equations for the first moments in the long-wavelength limit (expanded to quadratic order in $k$) are
\begin{equation}\label{linear_stability}
\partial_tm_i=\frac{k^2}{2}\bigg\{\sum_{j}\partial^2_i\beta V_e(R_{ij})(R_j-R_i)^2\bigg\}f(l,s;\alpha,\at)m_i,
\end{equation}
where
\begin{eqnarray}
f(l,s;\alpha,\at)&=& -D_0-s\kappa l^2+s^2\kappa l^4\alpha\nonumber\\
&-&s^3\kappa l^4\bigg\{\frac{\alpha^2}{\at}+\frac{1}{4\at}\sum_j\left[\partial_i^2\beta V_e(R_{ij})\right]^2\bigg\}.
\end{eqnarray}

Since $\partial_tm_i=r_k m_i$, the growth rate is proportional to $k^2f(l,s;\alpha,\at)$ up to $O(k^2)$.
It follows that a strictly uniform state ($k=0$) would not undergo small-amplitude dynamic instability regardless of the motor activity and susceptibility but the nonmoving state is barely stable in the absence of spatial modulation. Moreover, the sign of $f(l,s;\alpha,\at)$ determines the stability behavior for finite $k$. When the kick step size $l$ is small, $f$ is negative indicating diffusive relaxation towards the non-moving state. As $l$ increases, i.e., as the dynamics is more dominated by active processes, instabilities grow: $f$, thus $r_k$, becomes positive, signifying spontaneous initiation of a collective flow in a spatially modulated state, when $l$ exceeds a threshold value $l_\textrm{th}$ given by
\begin{equation}
l_\textrm{th}^2=\frac{1+\sqrt{1+4\frac{\alpha D_0}{\kappa}\left(1-s\big\{\frac{\alpha}{\at}+\frac{1}{4\alpha\at}\sum_j\left[\partial_i^2\beta V_e(R_{ij})\right]^2\big\}\right)}}{2(s\alpha)\left(1-s\big\{\frac{\alpha}{\at}
+\frac{1}{4\alpha\at}\sum_j\left[\partial_i^2\beta V_e(R_{ij})\right]^2\big\}\right)}.
\end{equation}
Consider the limit of high motor activity, such that $\at\gg\alpha$ and $\kappa\gg\alpha D_0$, we then have $l_\textrm{th}^2\sim1/(s\alpha)+D_0/(s\kappa)$, indicating that high motor susceptibility $s$ and kicking rate $\kappa$ as well as large mechanical gradient $\alpha$ lead to a low instability threshold in terms of kick step size (see Supplementary Fig.~S$1$ for a detailed illustration).
For the case of asymmetric susceptibility ($s_u\neq s_d$), the factor $(s_u-s_d)$ accompanies all the cubic-and-above odd powers in $\vec l$ in the expansion, leading to a lower threshold for the kick size compared to that for the symmetric case where deviations from effective equilibrium start at quartic order in $\vec l$.

We carried out the self-consistent calculation described earlier on a minimal model of the cytoskeleton as a cat's cradle \cite{CC} to determine ($\alpha,\at$). We now use this to obtain the growth rate.
As before \cite{JCP}, we study a model network consisting of nonlinear elastic filaments characterized by relaxed length $L_e$ and stretching stiffness $\beta\gamma$ built on a three dimensional random lattice of crosslinks at density $\rho$. The network connectivity $P_c$ is defined as the fraction of nearest-neighbor pairs of crosslinks connected by filaments. The unit of length is the average separation between the neighboring crosslinks.

\begin{figure}[htb]
\begin{center}
%\centerline{\includegraphics[angle=0, scale=0.75]{phase_diagrams.eps}}
\includegraphics[angle=0, scale=0.21]{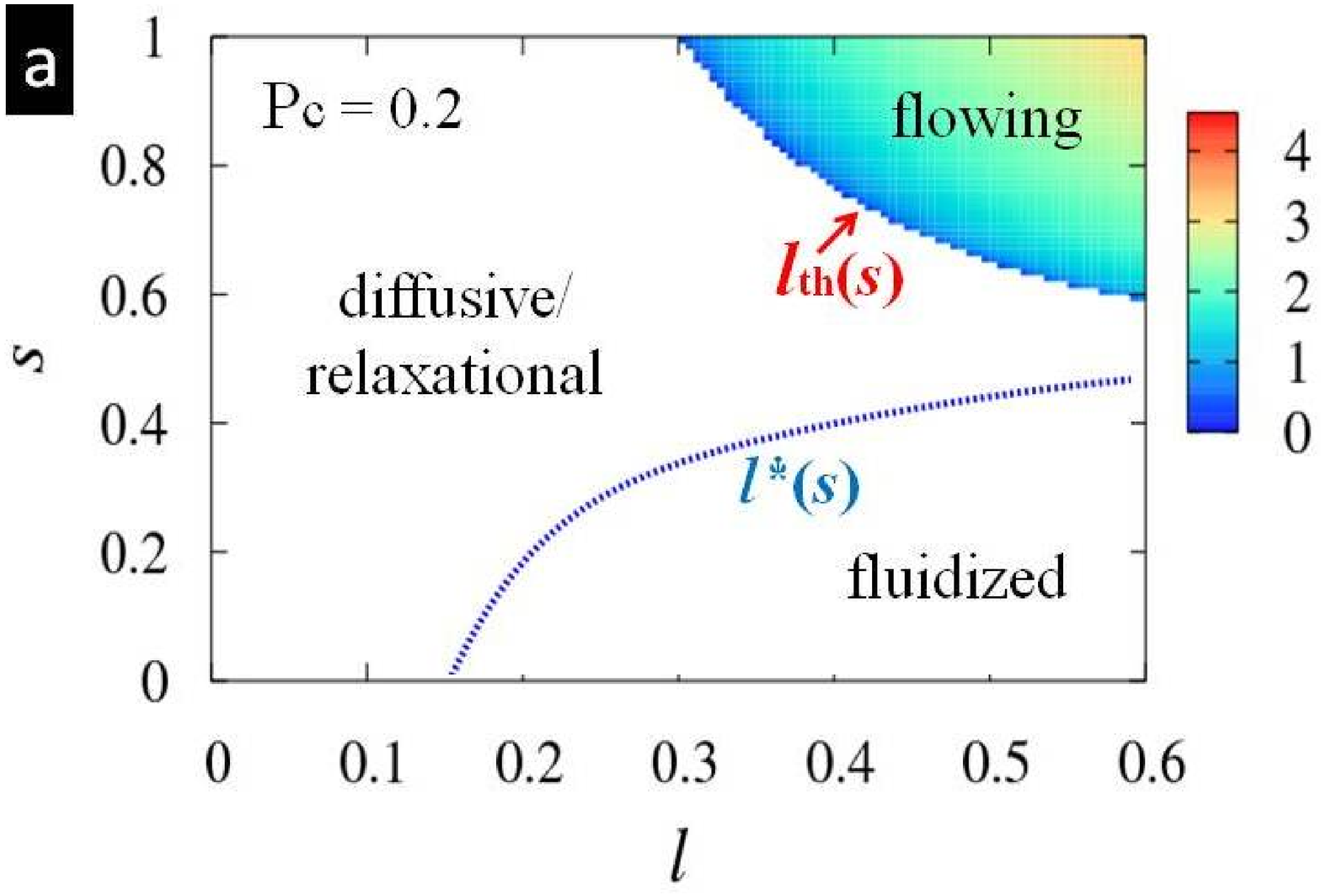}%
\includegraphics[angle=0, scale=0.21]{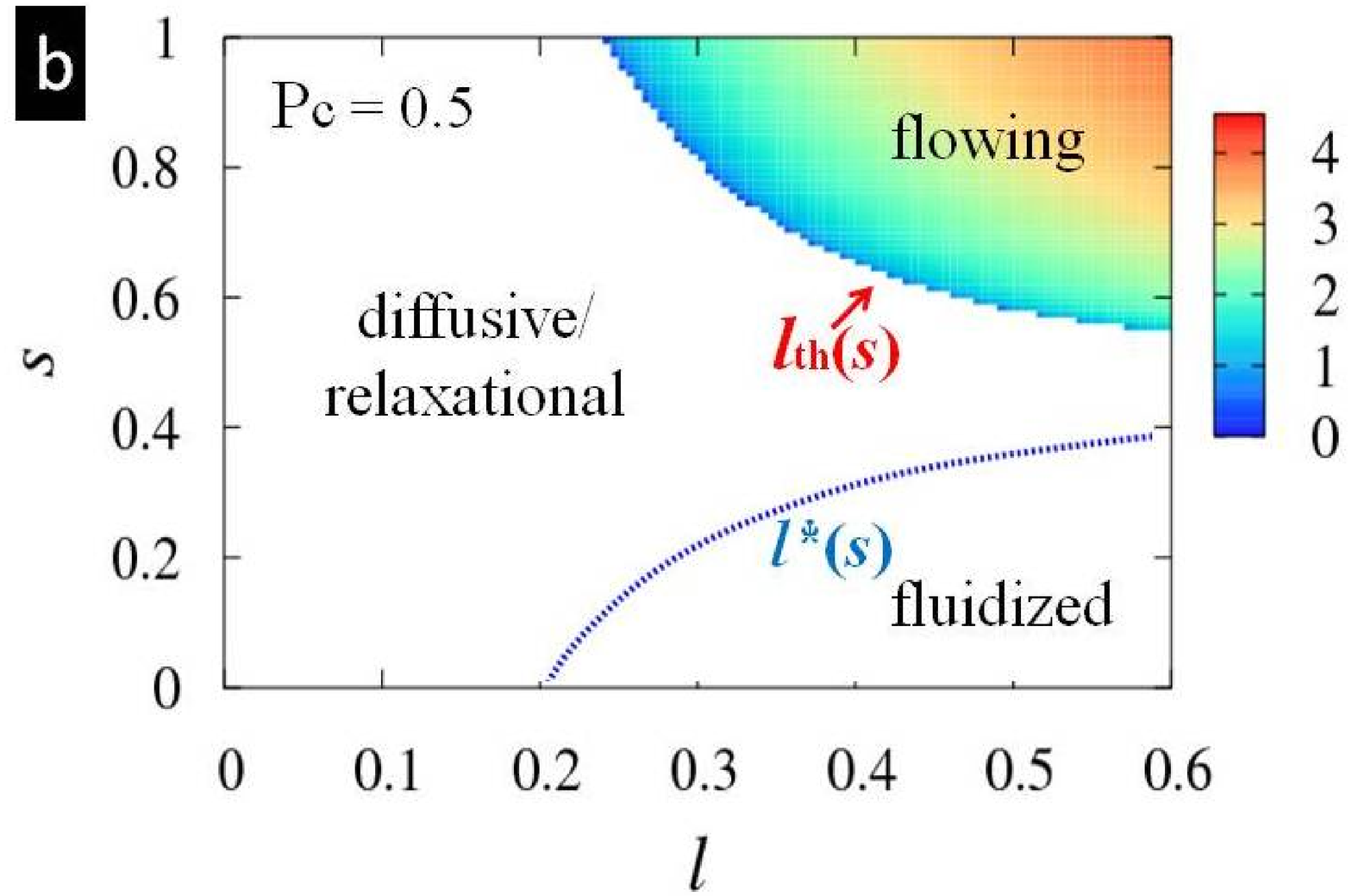}%
\includegraphics[angle=0, scale=0.21]{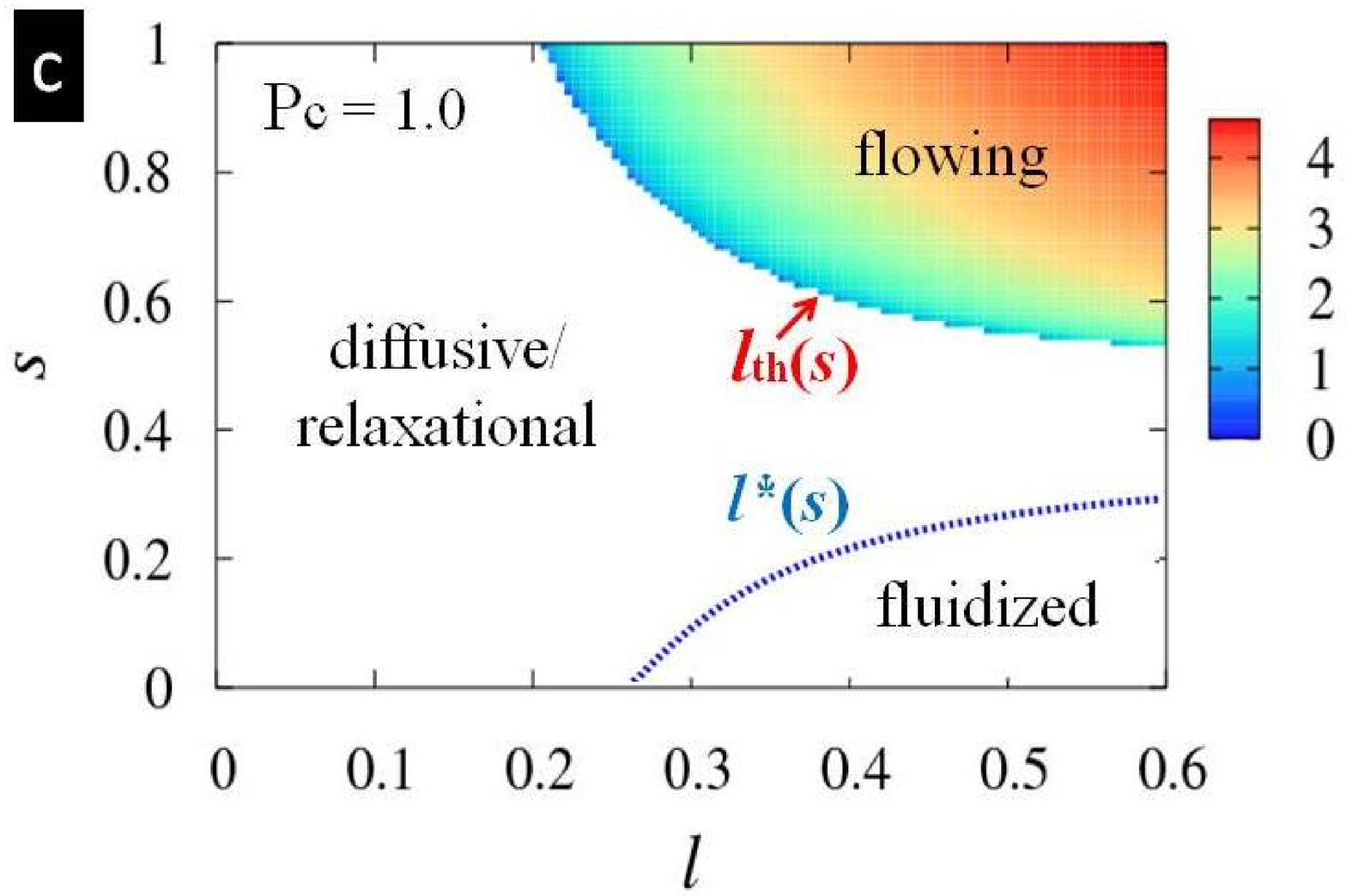}
\caption{Phase diagram for possible dynamic states. The parameter plane indicates the motor kick step size $l$ and the motor susceptibility $s$. The model cytoskeleton used to obtain $\alpha$ and $\at$ is characterized by crosslink density
$\rho=0.8$, relaxed length of the filaments $L_e=1.2$ and stretching stiffness $\beta\gamma=5$.
From panel (a) to (c), the network connectivity varies with $P_c=0.2, 0.5,$ and $1$, respectively.
$D_0=0.1$ and $\kappa=20$. In the flowing regime, there are stable nontrivial ($\alpha$,$\at$) solutions and positive $r_{k}$; in the diffusive/relaxational regime, while there are stable nontrivial ($\alpha$,$\at$) solutions there is a small negative $r_k$; in the fluidized regime, finite ($\alpha$,$\at$) solutions are unstable. As network connectivity rises, the flowing phase region expands whereas the fluidized state region shrinks. The logarithm of the normalized growth rate $r_k/k^2$ for flowing instability is color-coded, showing the increase of instability with $l$ and $s$.}
\label{diagram_vary_Pc}
\end{center}
\end{figure}

In Fig.~\ref{diagram_vary_Pc} we show the phase diagrams for possible dynamic states as a function of kick size $l$ and susceptibility $s$ for several values of the network connectivity. In all the cases, there are two stability boundaries, one for small $s$ ($s<1/2$), one for large $s$ ($s>1/2$).
In the low-s regime, as $l$ reaches a critical value, $l^*(s)$ (blue dotted line), finite solutions for ($\alpha, \at$) become unstable, the vanishing of the localization strength then suggests the system becomes fluidized.
In the high-$s$ corner, as the kick step size exceeds a threshold value $l_\textrm{th}$ (lower boundary of the color-coded region), an exponential growth of the first moments occurs for small but finite $k$ modes. This indicates the emergence of modulated flowing states. In this region, stable finite ($\alpha, \at$) solutions exist with $\at$ being considerably larger than $\alpha$, reflecting the enhancement of stability by susceptible motor kicking. Note that as motor susceptibility increases, the threshold kick step size decreases.
In the rest of the diagram, $\alpha$ and $\at$ are comparable and the negative growth rate indicates diffusive modes.
(Close to detailed balance, $s=1/2$, diffusive modes persist over the entire relevant range of $l$.)

In the figures, we color-code the logarithm of the normalized growth rate $r_k/k^2$ for the flow instability; the growth rate increases with kick size $l$ and susceptibility $s$.
Comparing the diagrams for different values of connectivity, we see that as $P_c$ increases, the region corresponding to the fluidized state shrinks, since increasing the number of bond constraints stabilizes the system against fluidization. On the other hand, the region corresponding to flow expands toward lower $l$ (and lower $s$ slightly),
suggesting that as the mechanical feedback enhances (larger $\alpha$ due to higher $P_c$), a smaller kick is able to trigger the flowing instability when the motors are susceptible.

To verify the predicted dynamic phases and visualize the structural development, we performed dynamic
Monte Carlo\cite{MC} simulations on the model cytoskeleton as a ``cat's cradle" \cite{CC, JCP}.
In these simulations we generated initially a three-dimensional random lattice of volumeless nodes
(mimicking the crosslinking proteins) and connected the nearest-neighbor nodes (defined by the first shell of
the pair distribution function) with nonlinear elastic bonds \cite{JCP} (mimicking the filamentous proteins)
at a given probability $P_c$ which characterizes network connectivity.
Thermal steps obey Brownian dynamics \cite{Brownian dynamics} whereas chemical moves follow the stochastic process defined by the model kicking statistics (equation \ref{kinetic rate}). We postpone for a more detailed article
describing all the simulations in different parameter regimes and only illustrate some qualitative features
hereafter. The simulations reveal an interesting interplay of network connectivity with the motor susceptibility dramatically affecting structural development.

At a relatively high network connectivity, $P_c\simeq0.5$ (average coordination number $z\simeq6$), force transmission through the bonds is efficient and the network structure remains statistically homogeneous in the presence of
the motor-driven processes. Nevertheless, varying motor susceptibility drastically changes the dynamics.

\begin{figure}[htb]
\begin{center}
%\centerline{\includegraphics[angle=0, scale=0.8]{dynamic_phases.eps}}
%\includegraphics[angle=0, scale=0.75]{dynamic_phases.eps}
\includegraphics[angle=0, scale=0.19]{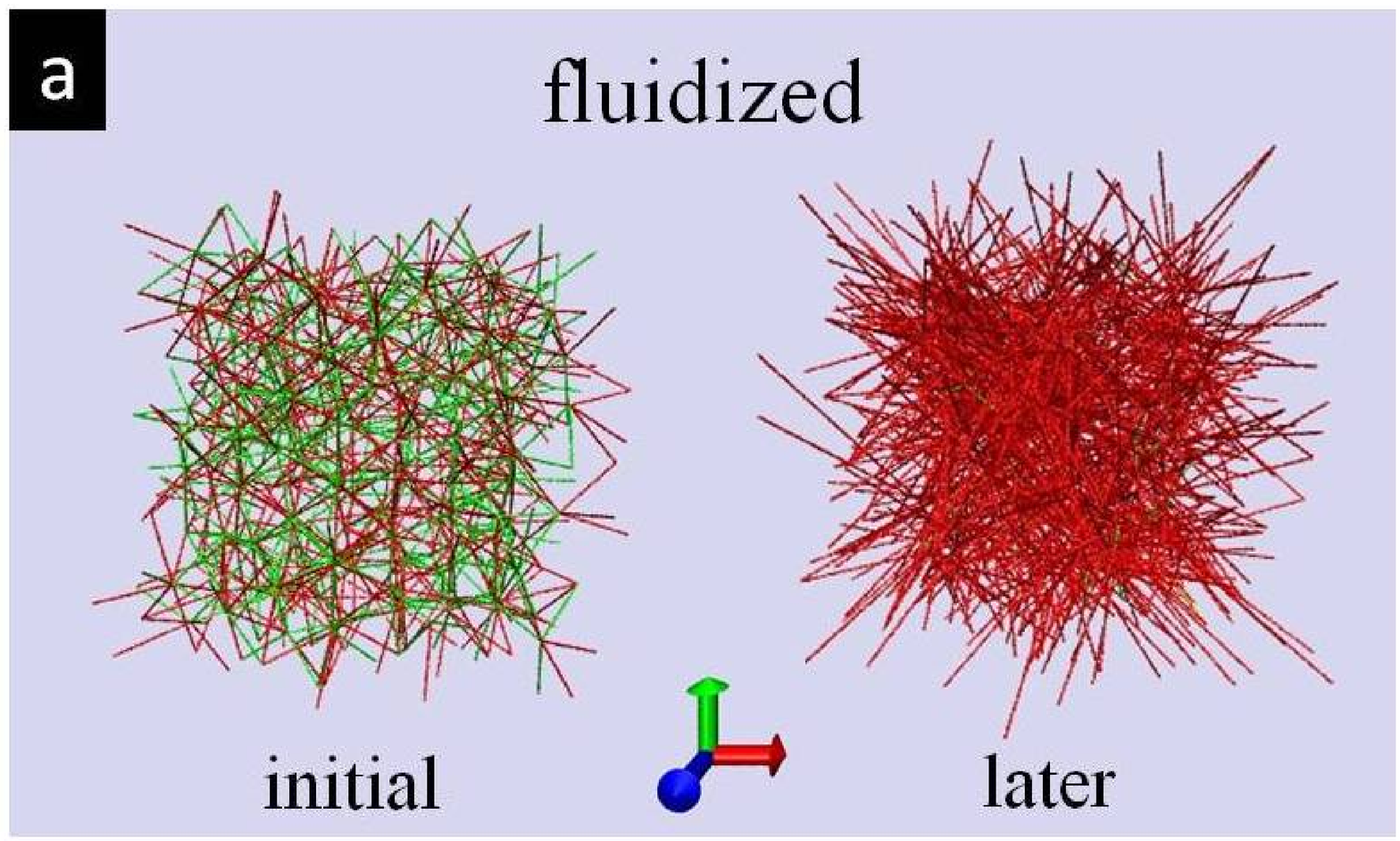}%
\includegraphics[angle=0, scale=0.19]{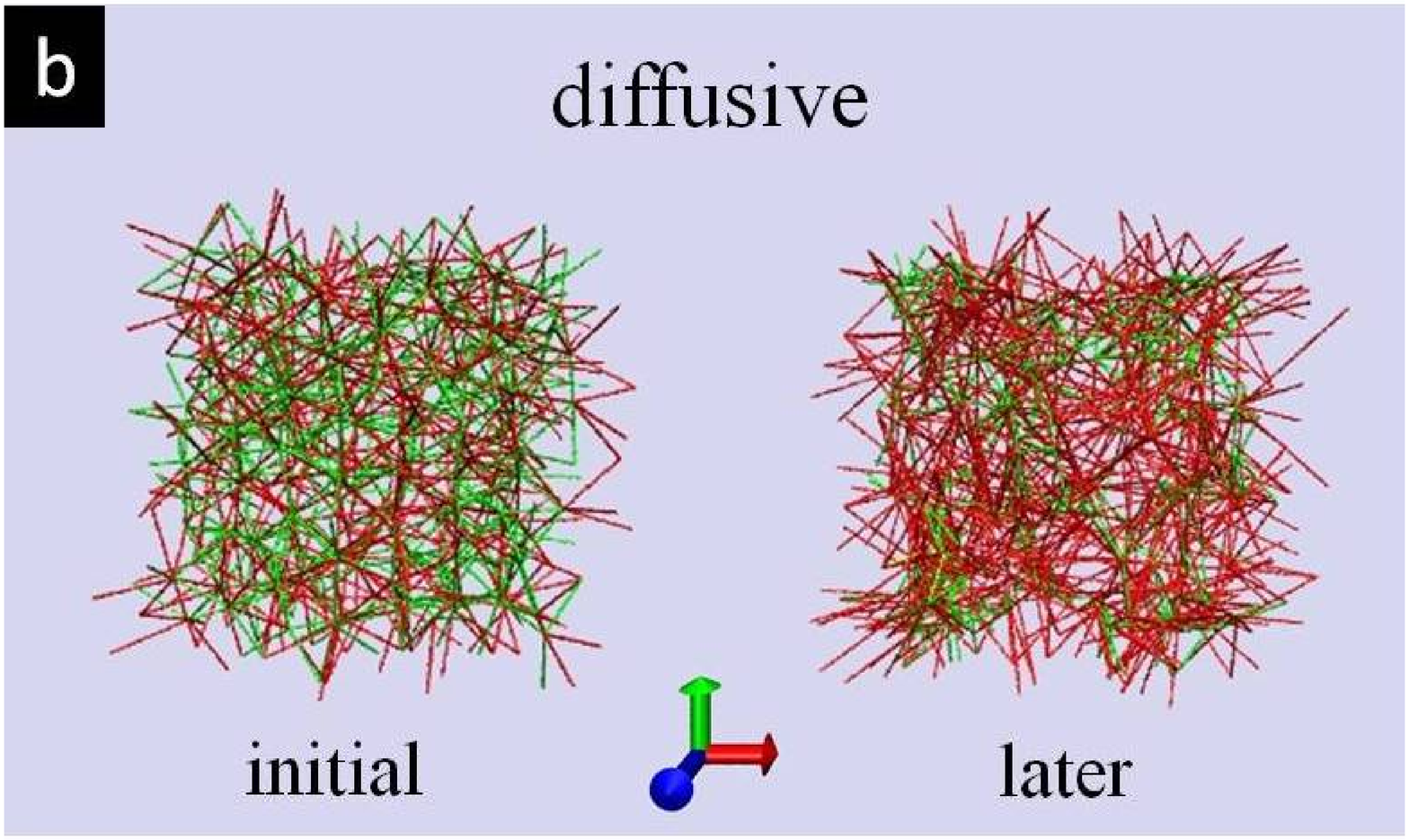}%
\includegraphics[angle=0, scale=0.19]{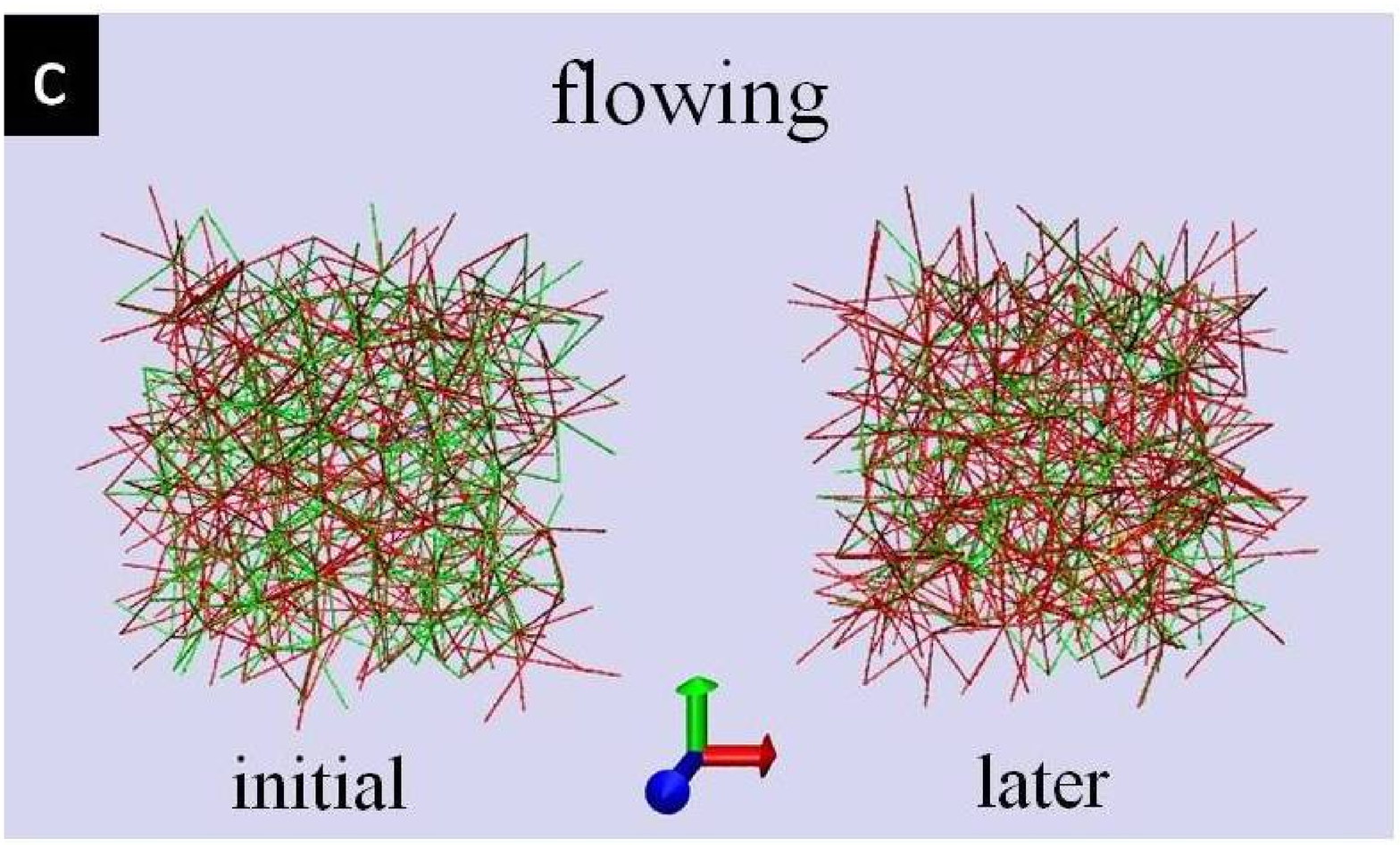}
\caption{Network structure for various dynamic phases.
(a) Fluidized phase with $s=0$. An initially relaxed (left) network rapidly tenses up (right)
under completely adamant motor kicking. Localization strength of the nodes vanishes and there is no net flow.
(b) Diffusive phase with $s=0.2$. Spots of concentrated tense/floppy filaments are visible. Nodes exhibit enhanced
diffusive motion with a finite localization strength. No spontaneous flow occurs.
(c) Flowing phase with $s=1$. Network structure remains homogeneous despite the spontaneous flowing motion, reflecting
the enhanced rigidity of the structure and coherence of motion by susceptible motor kicking.
$L_e=1.2$, $\beta\gamma=5$, $P_c=0.5$ and $l=0.25$.
Red lines stand for tense filaments and green lines for floppy filaments.
The small symbol at the bottom of each panel indicates the coordinate frame.}
\label{dynamic_phases}
\end{center}
\end{figure}

\textit{fluidized state}: Under completely adamant kicks ($s=0$) with a moderately large step size ($l>l^*$), nodes rapidly become fluidized.
Elastic stretching of the bonds imposes no constraint on the node motion resulting in vanishing localization strength
and zero net flow ($\vec{m}=0$). Consequently, as shown in Fig.~\ref{dynamic_phases}a, almost all the initially
floppy bonds (in green) get stretched (in red) and the network becomes very tense.

\textit{flowing state}: At the other extreme, however, under susceptible kicks ($s=1$) with above-threshold step size $l>l_\textrm{th}$,
a spontaneous and self-sustained flow develops and the nodes vibrate about a steadily moving fiducial lattice (see Supplementary Fig.~S$2$ for statistical characteristics of the flowing state).
We found spontaneous flow both for regular lattices and for random structures. A plausible physical mechanism is as follows: disorder in the structure, inherent in the quenched connectivity or dynamically generated
through initial random motions, gives rise to local force asymmetry
and thus local directed motion. Sufficiently large kicks then enhance an existing local force asymmetry and
trigger dynamic instability of the quiescent state; the resultant nucleation and propagation of local coordinated motion, mediated by force transmission and orchestrated by susceptible motor kicking, finally leads to a global concerted movement of the whole lattice. High motor susceptibility promotes cooperativity, resulting in ``rigidity" of the structure and coherence of the collective motion.
As can be seen in Fig.~\ref{dynamic_phases}c, even though the system undergoes a flow instability, the network structure remains homogenous without significant local distortions, reflecting the enhanced mechanical stability and
coherence of motion due to susceptible motor kicking.

\textit{diffusive state}: When the motors are only moderately susceptible to force ($s:0.2-0.5$) yet are not sufficiently cooperative to drive spontaneous flow, the system exhibits enhanced diffusive relaxation toward the effective equilibrium characterized by $T_\textrm{eff}$, leading to a homogeneous network structure with modest local density fluctuations.
The magnitude of density fluctuations and the tenseness of the network depend on the susceptibility.
At relatively low susceptibility ($s\leq0.3$), homogeneously distributed spots of concentrated tense or floppy filaments are visible (Fig.~\ref{dynamic_phases}b); as the susceptibility rises ($s\sim0.5$),
density fluctuations get weaker and the network becomes more homogeneous with a lower degree of stretching, closely resembling the flowing state (Fig.~\ref{dynamic_phases}c).
Therefore, for a homogeneous network at a relatively high connectivity the simulations verify the possible dynamic phases predicted by the analytical theory, that is for a sufficiently large kick step size, the system may undergo fluidized, diffusive and flowing states as the motor susceptibility increases.

\begin{figure}[htb]
%\begin{center}
\centerline{\includegraphics[angle=0, scale=0.37]{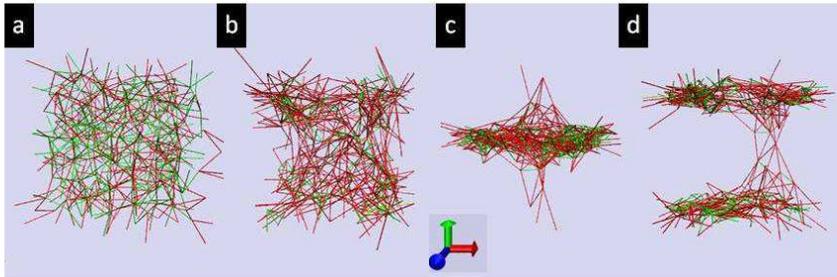}}
\caption{Temporal development of the phase separation.
At intermediate connectivity $P_c=0.3$ and under susceptible motor kicking ($s=1$) with a considerable step size $l=0.25$, the initially homogeneous system (a) phase separates into oscillating clumps and voids (b-d). Notice that the relaxed filaments (green) become concentrated within the planar clumps as well as the presence of highly stretched inter-clump filaments (red). $L_e=1.2, \beta\gamma=5$.}
\label{phase_separation}
%\end{center}
\end{figure}

Another noteworthy dynamical and structural development occurs at intermediate connectivity $P_c\simeq0.3$
($z\simeq3$--$4$). In this situation network connectivity is sufficient for tension percolation yet local force asymmetry becomes significant and widespread over the network. Under susceptible motor kicks with a considerable step size, dramatic spatial heterogeneity emerges and evolves into collapses and oscillations of the network in a particular spontaneously chosen spatial direction (Fig.~\ref{phase_separation} and Supplementary Movie SM$1$).
Apparently the overall tenseness of the structure is reduced by collapsing the network into clumps at the cost of a few highly stretched inter-clump filaments. Figures~\ref{phase_separation}c,d display the planar clumps where the floppy filaments (in green) concentrate and which are connected by highly stretched inter-clump bonds (in red).
Collapse can occur in various directions and the features of the spatial patterns are independent of
the system size.
Our previous analytical mean-field study of an equilibrium nonlinear-elastic network \cite{JCP} already suggests
the possibility of phase separation in this system at a finite effective temperature;
the pressure exhibits a non-monotonic dependence on the node concentration leading to mechanical instability
of homogeneous states. (Phase separation induced by contractile instability has also been predicted for active polar gels \cite{generic phase diagram}.)
When confined by boundaries, such as the cell membrane, these oscillating clumps may become
stationary wave patterns with a characteristic length scale of modulation, reminiscent of the mitotic spindles.

In the presence of susceptible motor kicking ($s\geq0.6$),
failure of force percolation at lower values of the network connectivity $P_c\leq0.2$\,($z<3$) also yields
phase separation but without a preferred direction of motion (Supplementary Movie SM$2$). Conversely when there are too many bond constraints at $P_c\geq0.6$ ($z>7$) there are significant mechanical barriers for motor dynamics which seems to slow down flow initiation and to reduce flow velocity.

\begin{figure}[htb]
%\begin{center}
%\centerline{\includegraphics[angle=0, scale=0.3]{f_vs_l_vary_alpha_s_kappa20.eps}}
\centerline{\includegraphics[angle=0,width=0.82\columnwidth]{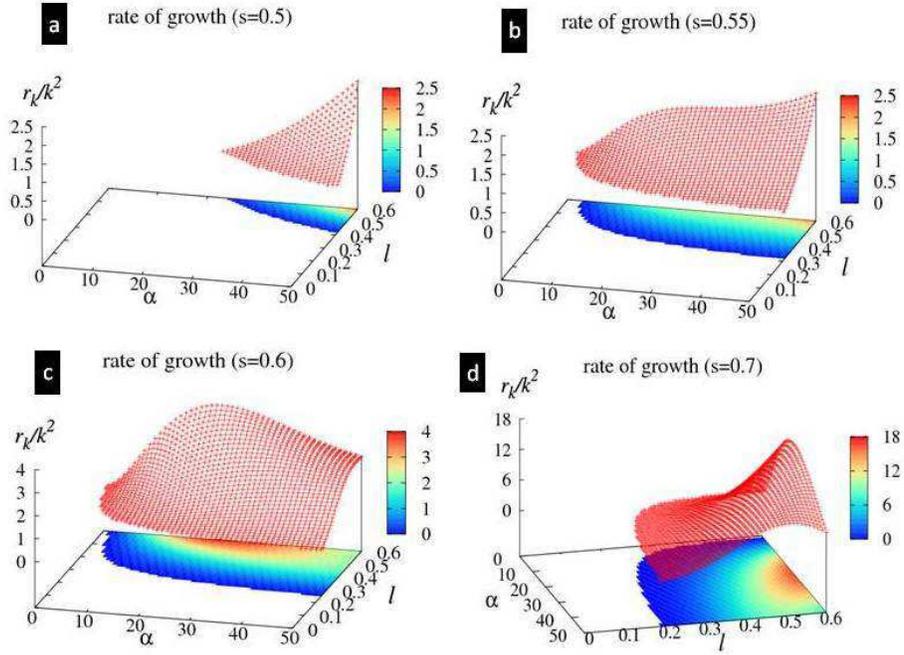}}
\caption{Dependence of growth rate upon localization strength at different motor susceptibility.
The $2$D surface and the contour map of the growth rate $r_k/k^2$ for flowing instability are displayed
over the parameter plane extended by the localization strength $\alpha$ of individual particles and
the motor kick step size $l$. $D_0=0.1$ and $\kappa=20$. (a) $s=0.5$; (b) $s=0.55$; (c) $s=0.6$; (d) $s=0.7$.
For susceptible motors with $s\geq0.6$, the growth rate develops a non-monotonic dependence on
the localization strength at a sufficiently large kick size, suggesting an optimal strength of
mechanical feedback for an efficient flowing motion.}
\label{rg_dep_alpha}
%\end{center}
\end{figure}

The analytic stability analysis leads to a similar connection between the growth rate of the dynamic instability to the number of bond constraints via the localization strength of individual nodes. In Fig.~\ref{rg_dep_alpha} we display the two dimensional surface as well as the contour map of the growth rate $r_k/k^2$ given by Eq.~(\ref{linear_stability}) in the flowing regime ($r_k>0$) as a function of localization strength $\alpha$
and kick size $l$ for a series of motor susceptibility $s$.
Close to detailed balance, i.e., $s=0.5$ (panel a), flowing instability emerges only at very high $\alpha$ and large $l$, and the growth rate increases with $\alpha$.
At $s=0.55$ (b), a plateau in the growth rate develops at relatively high $\alpha$ range.
For susceptible motors with $s\geq0.6$ (c and d), the growth rate exhibits a non-monotonic dependence on $\alpha$,
indicating that as the localization strength of individual constituents increases, growth of the flowing instability
first speeds up and then slows down; in other words, there exists an optimal localization strength
(or network connectivity) for the most efficient flow.
This is consistent with the numerical observation that in a tense network (high $\alpha$)
that exhibits spontaneous flowing motion driven by susceptible motor kicks ($s\geq0.6$),
as the network becomes increasingly connected, flow initiation as well as the flow velocity first increases
(for $P_c: 0.4\rightarrow0.6$ or $z:5\rightarrow7$)
but then decreases (for $P_c: 0.6\rightarrow1$ or $z:7\rightarrow12$); nevertheless a flowing instability persists
under these conditions (Supplementary Movie SM$3$-$7$).

\begin{figure}[htb]
%\begin{center}
\centerline{\includegraphics[angle=0, scale=0.24]{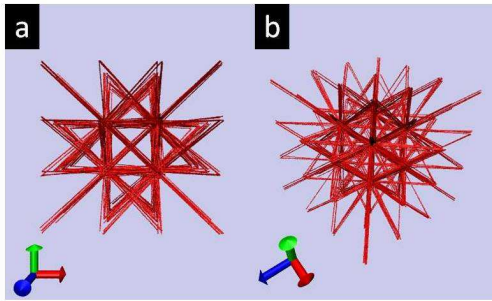}}
\caption{Aster-like patterns/Bundle-connected poles formed under kicks of uphill-prone motors with
a negative susceptibility. $s_u=-1,\,s_d=0$; $L_e=1.2$, $\beta\gamma=5$, $P_c=0.5$, and $l=0.25$.
(a) and (b) are snapshots of the system taken at the same instant from different view angles.}
\label{aster}
%\end{center}
\end{figure}

In principle motors could have slip bonds \cite{slip bonds_exp, slip bonds_th} such that an applied force shortens bond lifetimes by lowering the energy barrier rather than slowing up-hill moves. This slip-bond behavior leads to a negative motor susceptibility which in turn leads to a negative effective temperature. This implies an intrinsic thermodynamic instability. We investigated the course of structural development in this thermodynamically unusual situation. We consider the case for $s_u=-1, s_d=0$ where motors are insensitive to energetically downhill slope while they run faster when they go up against obstacles.
Starting with a disordered structure at a relatively high connectivity ($P_c=0.5$), the motorized network rapidly develops into a highly ordered and tense structure as shown in Fig.~\ref{aster}.

This interesting behavior is not hard to understand: due to the propensity to capture uphill slopes (negative $s_u$),
consistent with the negative effective temperature,
the motor kicks try to maximize the total energy by separating the bonded nodes as far as possible from each other;
within the periodic cubic box the nodes are therefore concentrated at the corners giving the maximal inter-node
separation and thus the highest degree of stretching. The resulting ``aster-like" patterns closely resemble
those formed by \textit{in vitro} reconstituted active gels \cite{novas of asters}, where unidirectional movement
of myosin II motors along the polar filament track toward the aster core (concentrated `plus' ends of actin filaments) results in considerable stress accumulation at the center, giving rise to the so-called ``novas of asters".
This is another demonstration of how motor susceptibility dramatically affects development of nonequilibrium structures.

In sum we have derived an analytical expression for the stability limits of quiescent active gels and
proposed a mechanism for spontaneous collective motion within a unified theoretical framework,
based on a systematic expansion of the motor-driven nonequilibrium dynamics in terms of the kick step size.
Simulations of a model cytoskeletal network further highlight that the interplay of network connectivity
with motor susceptibility dramatically affects the formation of nonequilibrium structures:
force percolation and mechano-chemical coupling conspire to drive and maintain the spontaneous flow,
whereas adamant motor kicks generally promote fluidization. Significant force imbalance sensed by susceptible motors induces phase separation into oscillating clumps. Uphill-prone motors with a negative susceptibility
give rise to a system at a negative effective temperature. Such motors drive the formation of
aster-like patterns resembling what is seen in reconstituted active gels.

We gratefully acknowledge the financial support from the Center for Theoretical Biological Physics
sponsored by the NSF via Grant PHY-0822283 and a critical reading of the manuscript by Olga K. Dudko.

{\bf Supplementary figures}

\begin{figure}[htb]
%\begin{center}
\centerline{\includegraphics[angle=0, scale=0.4]{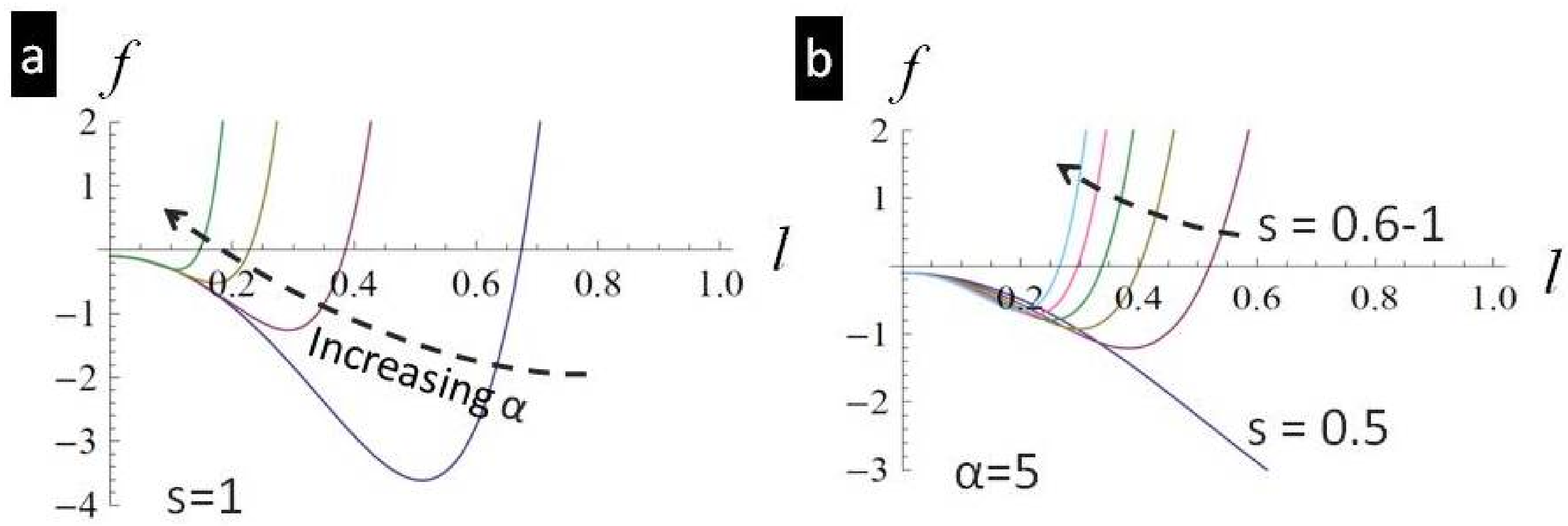}}
\label{f_vs_l_vary_alpha_s_kappa20}
%\end{center}
\end{figure}

\noindent
{\bf Fig.~S1 Dependence of threshold kick step size upon mechanical gradient $\alpha$ and motor susceptibility $s$.} We plot $f(l,s;\alpha,\tilde\alpha)$ given by equation (11) in the main text as a function of kick step size $l$ for $\kappa=20, D_0=0.1$. The crossing point of $f$-curve with $l$-axis locates $l_\textrm{th}$. (a) From right to left, $\alpha=0.1, 1, 10$ and $100$; $s=1$. $l_\textrm{th}$ decreases with increasing $\alpha$ value (dashed arrow), indicating that strong mechanical feedback facilitates instability onset. (b) For a moderate $\alpha$ value ($\alpha=5$), no instability onset occurs for $s\leq0.5$; only after $s$ reaches approximately $0.6$ does an instability emerge and increasing $s$ (dashed arrow) leads to lower $l_\textrm{th}$. Meanwhile, for a given $l$, $f$ increases with $s$, suggesting that more susceptible motors drive faster flows.\\

\begin{figure}[htb]
%\begin{center}
\centerline{\includegraphics[angle=-90,width=0.42\columnwidth]{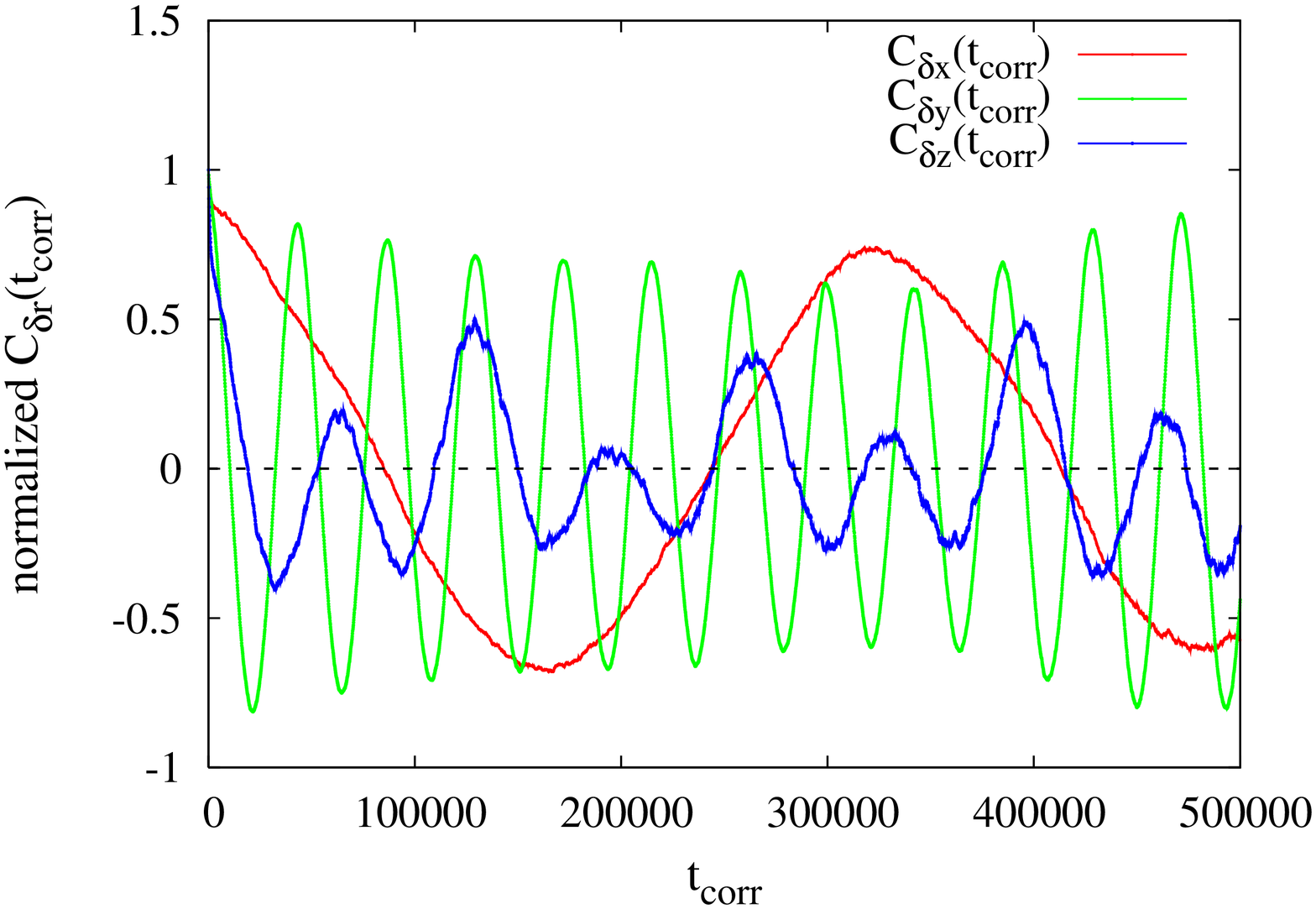}}
\label{correlation_measures}
%\end{center}
\end{figure}

\noindent
{\bf Fig.~S2 Statistical characteristics for the flowing state.}
We plot the ensemble-averaged temporal correlation of displacement fluctuations defined via the correlation function
$C_{\delta \vec r}(t_\textrm{corr})=\frac{1}{t_f-t_0}\int_{t_0}^{t_f}d\tau[\frac{1}{N}\sum_{i=1}^N\delta\vec{r}_i(\tau)]
[\frac{1}{N}\sum_{j=1}^N\delta\vec{r}_j(\tau+t_\textrm{corr})]$,
where $\delta\vec{r}_i(t)=\vec{r}_i(t)-\langle\vec{r}_i(t)\rangle$ with the bracket indicating an average over a steady-state time window. The integral averages the starting time $\tau$ over a wide time range ($t_0$, $t_f$).
Correlations in three orthogonal spatial directions (x: red; y: green; z: blue) all exhibit periodic oscillations in time (in Monte Carlo units) reflecting the spontaneous and coherent flowing motion across the periodic simulation box. Other correlation measures exhibit consistent oscillatory patterns.
$L_e=1.2$, $\beta\gamma=5$, $P_c=0.5$; $l=0.25, s=1$.

\end{document}